 \documentclass[floatfix,fceqn]{elsart}
\usepackage{amsmath,amssymb,graphicx,epstopdf,placeins,float}
\usepackage{diagbox}
\usepackage{pstricks}
\usepackage[numbers,sort&compress]{natbib}
\usepackage{hyperref}
 
 \newcommand{\beq}[1]{\begin{equation}\label{#1}}
 \newcommand{\eeq}{\end{equation}}
 \newcommand{\beg}[1]{\begin{gather}\label{#1}}
 \newcommand{\eeg}{\end{gather}}
 \newcommand{\bea}[1]{\begin{eqnarray}\label{#1}}
 \newcommand{\eea}{\end{eqnarray}}

 \newcommand{\m}{\rule[1.5pt]{2pt}{0.5pt}}
 
 \newcommand{\rhoONEdistribute}[1]{{\psset{unit=#1}
 \begin{pspicture}(-0.1,0)(0.6,0.35)
 \psline[linewidth=0.5pt]{->}(-0.1,0)(0.5,0)
 \psline[linewidth=0.5pt](0,-0.05)(0,0.35)
 \psdots[dotstyle=square](0.06,0.06)(0.06,0.16)(0.06,0.26)
 \end{pspicture}
 }}
 \newcommand{\rhoTWOdistribute}[1]{{\psset{unit=#1}
 \begin{pspicture}(-0.1,0)(0.6,0.35)
 \psline[linewidth=0.5pt]{->}(-0.1,0)(0.5,0)
 \psline[linewidth=0.5pt](0,-0.05)(0,0.4)
 \psdots[dotstyle=square](0.16,0.06)(0.16,0.16)(0.16,0.26)
 \end{pspicture}
 }}
 \newcommand{\rhoTHREEdistribute}[1]{{\psset{unit=#1}
 \begin{pspicture}(-0.1,0)(0.6,0.35)
 \psline[linewidth=0.5pt]{->}(-0.1,0)(0.5,0)
 \psline[linewidth=0.5pt](0,-0.05)(0,0.4)
 \psdots[dotstyle=square](0.26,0.06)(0.26,0.16)(0.26,0.26)
 \end{pspicture}
 }}
 \newcommand{\rhoFOURdistribute}[1]{{\psset{unit=#1}
 \begin{pspicture}(-0.1,0)(0.6,0.35)
 \psline[linewidth=0.5pt]{->}(-0.1,0)(0.5,0)
 \psline[linewidth=0.5pt](0,-0.05)(0,0.4)
 \psdots[dotstyle=square](0.06,0.06)(0.06,0.16)(0.16,0.06)
 \end{pspicture}
 }}
 \newcommand{\rhoFIVEdistribute}[1]{{\psset{unit=#1}
 \begin{pspicture}(-0.1,0)(0.6,0.35)
 \psline[linewidth=0.5pt]{->}(-0.1,0)(0.5,0)
 \psline[linewidth=0.5pt](0,-0.05)(0,0.4)
 \psdots[dotstyle=square](0.06,0.06)(0.16,0.16)(0.16,0.06)
 \end{pspicture}
 }}
 \newcommand{\rhoSIXdistribute}[1]{{\psset{unit=#1}
 \begin{pspicture}(-0.1,0)(0.6,0.35)
 \psline[linewidth=0.5pt]{->}(-0.1,0)(0.5,0)
 \psline[linewidth=0.5pt](0,-0.05)(0,0.4)
 \psdots[dotstyle=square](0.16,0.06)(0.16,0.16)(0.26,0.06)
 \end{pspicture}
 }}
 \newcommand{\rhoSEVENdistribute}[1]{{\psset{unit=#1}
 \begin{pspicture}(-0.1,0)(0.6,0.35)
 \psline[linewidth=0.5pt]{->}(-0.1,0)(0.5,0)
 \psline[linewidth=0.5pt](0,-0.05)(0,0.4)
 \psdots[dotstyle=square](0.16,0.06)(0.26,0.16)(0.26,0.06)
 \end{pspicture}
 }}
 \newcommand{\rhoEIGHTdistribute}[1]{{\psset{unit=#1}
 \begin{pspicture}(-0.1,0)(0.6,0.35)
 \psline[linewidth=0.5pt]{->}(-0.1,0)(0.5,0)
 \psline[linewidth=0.5pt](0,-0.05)(0,0.4)
 \psdots[dotstyle=square](0.06,0.06)(0.06,0.16)(0.26,0.06)
 \end{pspicture}
 }}
 \newcommand{\rhoNINEdistribute}[1]{{\psset{unit=#1}
 \begin{pspicture}(-0.1,0)(0.6,0.35)
 \psline[linewidth=0.5pt]{->}(-0.1,0)(0.5,0)
 \psline[linewidth=0.5pt](0,-0.05)(0,0.4)
 \psdots[dotstyle=square](0.06,0.06)(0.26,0.16)(0.26,0.06)
 \end{pspicture}
 }}
 \newcommand{\rhoTENdistribute}[1]{{\psset{unit=#1}
 \begin{pspicture}(-0.1,0)(0.6,0.35)
 \psline[linewidth=0.5pt]{->}(-0.1,0)(0.5,0)
 \psline[linewidth=0.5pt](0,-0.05)(0,0.4)
 \psdots[dotstyle=square](0.06,0.06)(0.16,0.06)(0.26,0.06)
 \end{pspicture}
 }}

\begin{document}

\begin{frontmatter}
\title{Resolving the Schwarzschild singularity in both classic and quantum gravity}
\author[bjut,cas]{Ding-fang Zeng}
\ead{dfzeng@bjut.edu.cn}
\address[bjut]{Theoretical Physics Division, College of Applied Sciences, Beijing University of Technology, China 100124}   
\address[cas]{State Key Laboratory of Theoretical Physics, Institute of Theoretical Physics, Chinese Academy of Sciences\\Beijing, China 100124}
\date{February, 19th 2017}
\begin{abstract}
The Schwarzschild singularity's resolution has key values in cracking the key mysteries related with black holes, the origin of their horizon entropy and the information missing puzzle involved in their evaporations. We provide in this work the general dynamic inner metric of collapsing stars with horizons and with non-trivial radial mass distributions. We find that static central singularities are not the final state of the system. Instead, the final state of the system is a periodically zero-cross breathing ball. Through 3+1 decomposed general relativity and its quantum formulation, we establish a functional Schr\"odinger equation controlling the micro-state of this breathing ball and show that, the system configuration with all the matter concentrating on the central point is not the unique eigen-energy-density solution. Using a Bohr-Sommerfield like ``orbital'' quantisation assumption, we show that for each black hole of horizon radius $r_h$, there are about $e^{r_h^2/\ell^2_\mathrm{pl}}$ allowable eigen-energy-density profile. This naturally leads to physic interpretations for the micro-origin of horizon entropy, as well as solutions to the information missing puzzle involved in Hawking radiations. 
\end{abstract}


\end{frontmatter}
{\it\bfseries Motivation and logic.} Although string theory and loop gravity \cite{strominger1996,mathur2005,lqgEntropy1996,lqgEntropy1997} both give interpretations 
for the microscopic origin of some --- loop claims any --- black holes' entropy \cite{Wald1999}, partly due to lacks of a common semi-classic picture, none of them is considered the final answer \cite{LectureNotesPolchinski}. Related with the physic of micro-states, is the black hole's information missing puzzle. That is, when a black hole evaporates, where does the information it carries go away  \cite{hawking1976,fireworksAMPS2012,fireworksAMPS2013,fireworksCOP2016,hawking1509}?
In principles, any interpretation for the micro-states of a black hole should also tell us how they changes when it evaporates. In practices, almost all existing resolutions\cite{Preskill1992,Giddings1995,Hartle1997,Giddings1998,Nikolic2009,hawking1401,Nikolic201505} to this puzzle are regardless of the quantum theories' micro-state interpretation. Very recently, Hawking, Perry and Strominger \cite{Hawking2016} propose to solve this question+puzzle in a unifying framework of infinite number of hidden symmetries. Their proposal is still in completion but seems very hard to be dis/verified observationally. The purpose of this work is to provide a simple but dis/verifiable semi-classic picture, as well as quantisation method for the micro-state of black holes and the corresponding resolutions to the information missing puzzle involved in Hawking radiations. The core of the work is the Schwarzschild singularity's resolution.

Our logic is, if \footnote{We will provide exact classic solution examples displaying that the final state is indeed a zero-cross breathing ball, thus no contradictions with Penrose and Hawking's singularity theorem occurs here.} central singularities are not the final fate of collapsing stars, then all question+puzzles related with the micro-state of black holes must be understandable from inner structures of the collapsing star leading to its formation. Obviously, for a Schwarzschild black hole, the most natural micro-structure inheritable from its parental star is the radial mass distribution $m(0,r)$ and evolving speed $\dot{m}(0,r)$ at some initial epochs. Non-radial local random motions inside an externally-looking spherical symmetric star are although possible, due to the fact that $n\propto m_\mathrm{total}\propto r_h$, i.e. the particle number linearly depends on the mass thus on the horizon size of the black hole, they contribute to the entropy of the system only of $\mathcal{O}[r_h]$, obviously negligible relative to the horizon entropy $\mathcal{O}[r_h^{n-1}]$ in $n+1$ dimensional space-times. We will show that in the quantum formulation of 3+1 decomposed general relativities, the micro-states of the collapsing star is defined by eigen-energy-density solutions of a functional Schr\"odinger equation.  For very large this kind of star, through a Bohr-Sommerfield like ``orbital'' quantisation assumption, we show that the degeneracy of eigen-solutions is about $e^{r_h^2/r^2_\mathrm{pl}}$. Since each of these degenerating stars has its own characteristic de-horizon/expansion speed determined by its radial mass distribution and could be measured as its identifying accordance, no information will be missed during a black hole's evaporation. 

The content of this work is organised as follows. The next section will focus on classic metric exploration of collapsing stars with general radial mass distributions. While the next next section provides quantum descriptions for the physic pictures uncovered in section II. We then cost two sections discussing the micro-states' number counting of black holes and the resolution of information missing puzzle involved in Hawking radiations. The last section is our conclusion and prospects for future works.

{\it\bfseries Inner structure of black holes, classic picture.} Historically, Oppenhemer and Snyder (OS) \cite{oppenheimer1939,MTWbook} are the earliest physicists to consider the inner structure of Schwarschild black holes. But they assumes that matter contents inside the horizon is uniformly distributed thus excludes the possibility of non-unique micro-states. Yodzis, Seifert and M\"{u}ller (YSM) \cite{Yodzis1973,Yodzis1974} considered layering matter contents inside horizons in constructing counter examples to the cosmic census hypothesis. But they noted nothing about this layering structure with the micro-state of black holes. In both OS and YSM's works, inner metrics of the black holes were written in co-moving spatial coordinates, which due to shell-crossing phenomenas will become invalid before central singularities formation thus of no use in quantum resolutions of the singularity. As comparisons, our metrics in this work uses only Schwarzschild-like spatial coordinates \cite{Vaidya1950,Vaidya1970}. They are thus valid during the whole process of the central singularity's formation.

We find that full geometries of a collapsing star with general radial mass distributions could be written as
\begin{gather}
ds^2=-h^{-1}A(\tau,r)d\tau^2+h^{-1}dr^2+r^2d\Omega_2^2
\label{metricInternal}
\\
h=1-\frac{2m(\tau,r)}{r},~r<r_0\rule{25mm}{0pt}
\nonumber
\\
A=\frac{\dot{m}^2}{{m'}^2}+h\rule{48mm}{0pt}
\nonumber
\end{gather}
where $r_0$ is the initial radius of the dust star and $m(\tau,r)$, the mass of all contents inside the sphere of radius $r$ at time $\tau$, with $\tau$ being the proper time of freely collapsing matter contents. 
To connect with the Schwarzschild metric on the boundary of the star, it is required that
\beq{}
A(0,r_0)=1,~d\tau =hdt
\eeq
On there, $\tau$ happens to be the proper time of freely falling observers in the Schwarzschild background, whose equations of motion just read $h\dot{t}=1$, $\dot{r}^2=1-h$.
Inside the collapsing star, those observers will co-move with the matter contents are thus controlled by $u^0=1$, $u^1=-\frac{\dot{m}}{m'}$ and Einstein equations $R_{\mu\nu}-\frac{1}{2}g_{\mu\nu}R=\rho u_{\mu}u_{\nu}$ in the zero-pressure dust star case
\beq{}
\frac{m''}{m'}=\frac{\dot{m}'}{\dot{m}}
-\frac{m}{r^2}\frac{{m'}^2}{\dot{m}^2}
-\frac{2m}{r(r-2m)}
\label{eqConstraint}
\eeq
\beq{}
\frac{\ddot{m}}{\dot{m}}\frac{m'}{\dot{m}}=\frac{\dot{m}'}{\dot{m}}+\frac{2m}{r^2}\frac{m'^2}{\dot{m}^2}
+\frac{2m}{r(r-2m)}
\label{eqEvolution}
\eeq
It should be noted that the metric ansatz \eqref{metricInternal} is valid regardless matter contents consisting the collapsing star has pressures or not. However, zero-pressure condition enters equations \eqref{eqConstraint} and \eqref{eqEvolution}. They are thus valid only for zero-pressure dust stars. 

Equations \eqref{eqConstraint}-\eqref{eqEvolution} being valid simultaneously implies a redundancy, i.e. two equations controlling one variable $m(\tau,r)$'s evolution. We only need $m(0,r)$ instead of $\{m(0,r),\dot{m}(0,r)\}$ as a whole to specify initial status of the system. This is a general feature of Einstein equation. Similar things also occur in cosmologies. There the evolution of the homogeneous and isotropic universe is controlled by two Friedmann equations
\beq{}
\frac{\dot{a}^2}{a^2}+\frac{k}{a^2}=\frac{1}{3}\big(\rho+\Lambda\big)
\label{friedmannA}
\eeq
\beq{}
2\frac{\ddot{a}}{a}+\frac{\dot{a}^2}{a^2}+\frac{k}{a^2}=-(\rho-\Lambda)
\label{friedmannB}
\eeq
The first is time-first-order while the second is time-second-order. Obviously we need only know the initial value of $a(0)$ to predict its future evolutions. 

\begin{figure}[h]\begin{center}
\includegraphics[scale=0.47]{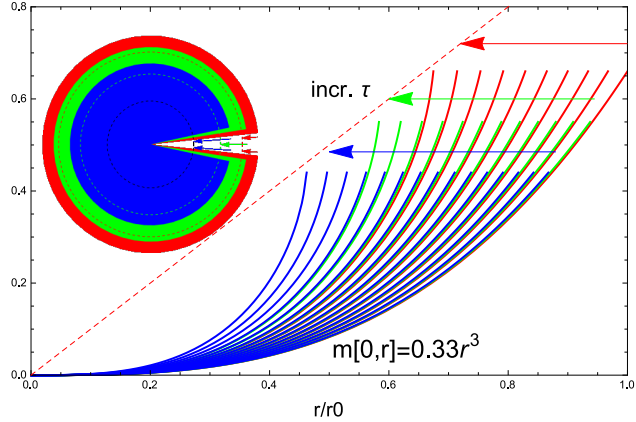}
\end{center}
\caption{Red lines display the variation of mass distributions inside the whole star $r<1.00r_0$; green lines, the variation of mass distributions inside the sphere $r<0.94r_0$; while the blue ones, that inside the sphere of $r<0.87r_0$. Along the red dashed line, $1-\frac{2m}{r}=0$. The initial distribution is assumed $m(0,r)=c\cdot\mathrm{min}\{r,r_0\}^3$. Changing parameters $c$, $q$ and $r_0$ will not change this picture qualitatively. But more general initials like $m(0,r)=\sum_{i}c_i\mathrm{min}\{r,r_{i}\}^{q_i}$ could lead to collapsing stars with multi-horizons among which some may be finished forming earlier than the outmost one.}
\label{figmtrq3}
\end{figure}

Due to redundancies in the equation of motion, evolutions of a collapsing star are completely determined by its initial mass distribution $m(0,r)$. For example, for the following non-singular, no-horizon initial distributions,
\begin{gather}
m(0,r)=c\cdot r^q,~0<1-\frac{2c\cdot r^q}{r},~0<r<r_0
\label{m0condition}
\\
m(0,r)=cr_0^q,~r_0<r, \rule{39mm}{0pt}
\nonumber
\end{gather}

the corresponding $\dot{m}(0,r)$ is non-freely settable, it is determined by the constraint \eqref{eqConstraint} ($\pm$ correspond collapsing/expanding respectively)
\begin{gather}
\dot{m}(0,r)=\pm r^{q-1}\Big[\frac{b-\frac{c^2q^2}{q+1}(1-2cr^{q-1})^{\frac{q+1}{q-1}}}{(1-2cr^{q-1})^\frac{2}{q-1}}\Big]^\frac{1}{2}
\label{dm0condition}
\\
\dot{m}(0,r)=0,~r_0<r\rule{42mm}{0pt}
\nonumber
\end{gather}
In these formulas, $m(0,r)$ is the initial mass distribution, $r_0$, $cr_0^q$ and $q(>\!\!1)$ are the initial star radius, total mass and pattern parameter of distributions respectively. Obviously, more general initials could be implemented by superpositions of the form $m(0,r)=\sum_{i}c_i\mathrm{min}\{r,r_{i}\}^{q_i}$. With initial conditions \eqref{m0condition} + \eqref{dm0condition} as a concrete example, 
second order forward Runge-Kuta algorithm could be used to integrate equations \eqref{eqEvolution} and \eqref{eqConstraint} simultaneously. We displayed the results in FIG. \ref{figmtrq3}.

From FIG.\ref{figmtrq3}, we firstly note that no matter how the initial distribution is, near the outmost horizon entrance point, the mass function has linear-inversely divergent first order derivative $m'(\tau,r\rightarrow r_h)\propto(r-r_h)^{-1}$, so that
\beq{}
m(\tau,r)\xrightarrow{2m\rightarrow r_h}a\log(1-r/r_h)+b
\label{mfLGsingular}
\eeq
This will play key roles in our derivations of the black hole entropy's area law and can be seen from the limit analysis of equations \eqref{eqConstraint} and \eqref{eqEvolution} directly, in which $\frac{m'}{\dot{m}}=(\frac{dr}{d\tau})^{-1}$ is finite, but $\frac{m''}{m'}$, $\frac{\ddot{m}}{\dot{m}}$, $\frac{\dot{m}'}{\dot{m}}$ and $\frac{2m}{r-2m}$ are all linear-inversely divergent. This forms a technique firewall prohibiting us from evolving the differential system beyond the horizon formation epoch. However, if we calculate the physical mass/energy density
\beq{}
\rho=-T^0_{~0}-T^1_{~1}=\frac{2(r-2m){m'}^3}{r^2[(r-2m){m'^2}+r\dot{m}^2]}
\label{rhoExpression}
\eeq
we will find that the result is everywhere regular at this epoch. So this firewall is not a wall of infinite physical energy density. We guess they may be related the AMPS \cite{fireworksAMPS2012,fireworksAMPS2013} firewalls techniquely.

\begin{figure}[h]
\begin{center}\includegraphics[scale=0.7, clip=true, bb=0 30 374 207]{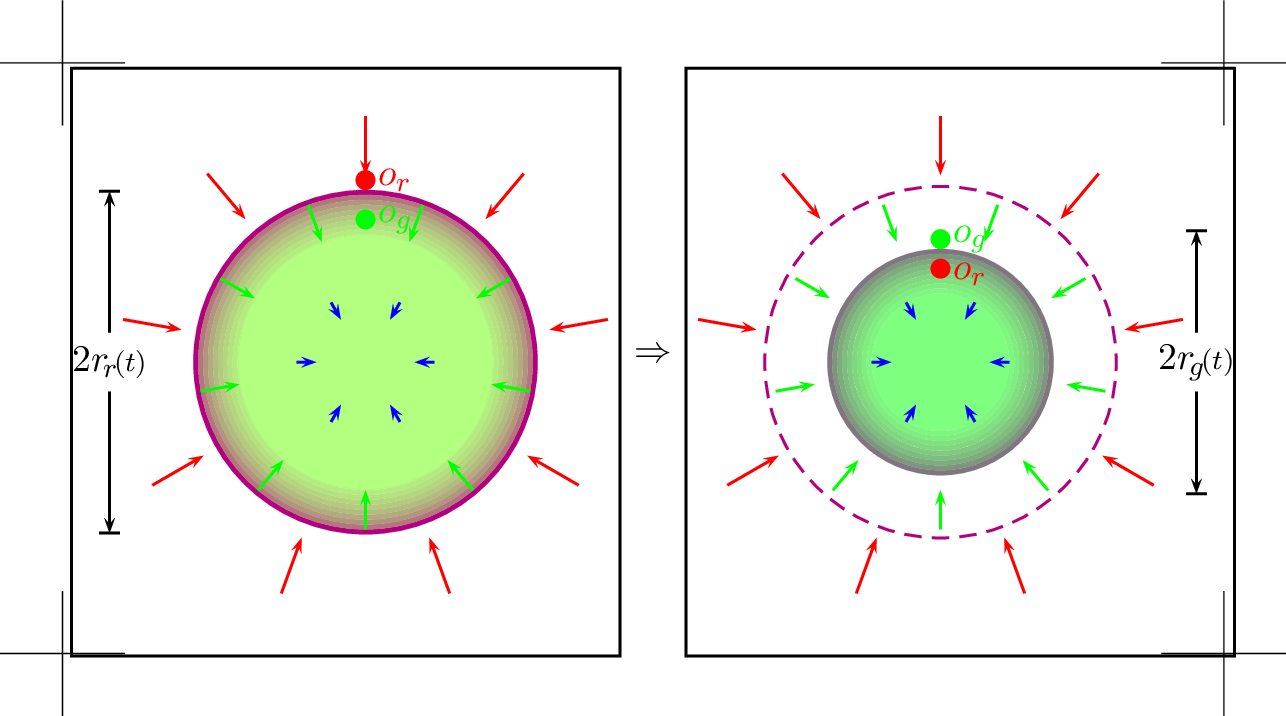}\end{center}
\caption{$r_{\!r}(t)$ and $r_{\!g}(t)$ are radial coordinates of two observers co-moving with a collapsing star, $r_{r0}$, $r_{g0}$ are their initial values. Newton mechanics tell us that $\ddot{r}_i=-\frac{G\rho_{i0}r^3_{i0}}{r^2_i}$ $\Rightarrow$ $r_i=r_{i0}(1-\frac{t}{t_{i0}})^\frac{2}{3}$, with $t_{i0}^2=\frac{1}{G\rho_{i0}}$ denoting the time observer i falling to the central point. Obviously, if the average density $\rho_{r0}$ of masses inside the sphere $r_{r0}$ is larger than that inside $r_{g0}$, then the observer r will fall earlier than g to the center of the star, during which shell crossing happens somewhere inside sphere $r_{g0}$.}
\label{figShCrossing}
\end{figure}
The second point we can see from FIG.\ref{figmtrq3} is that, although the collapsing star as a whole will quickly contract into its horizon surface, its inner sub-star will not do so! The more inner sub-star needs more longer time to contract into their own horizon surface. The most inner sub-star almost needs infinite length of time to fall onto the central point. Combining this fact with Penrose and Hawking's singularity theorem  \cite{Penrose,geroch1979,LargeScaleStructure} which says that any of this collapsing stars will collapse to the central point in finite proper times, we infer that during the outmost mass-shell's collapsing to the central point, it must shell-cross all mass-shells initially more close to the central point, see FIG.\ref{figShCrossing} for pictures. This shell-crossing phenomena was firstly mentioned by YSM in Ref.\cite{Yodzis1973,Yodzis1974} as the origin of naked singularity thus counter examples to the cosmic consensus hypothesis. For this reason, its physic value is long-termly ignored, or even negatively viewed. 

\begin{figure*}[ht]
\begin{center}\includegraphics[scale=0.85,clip=true, bb=0 30 516 300]{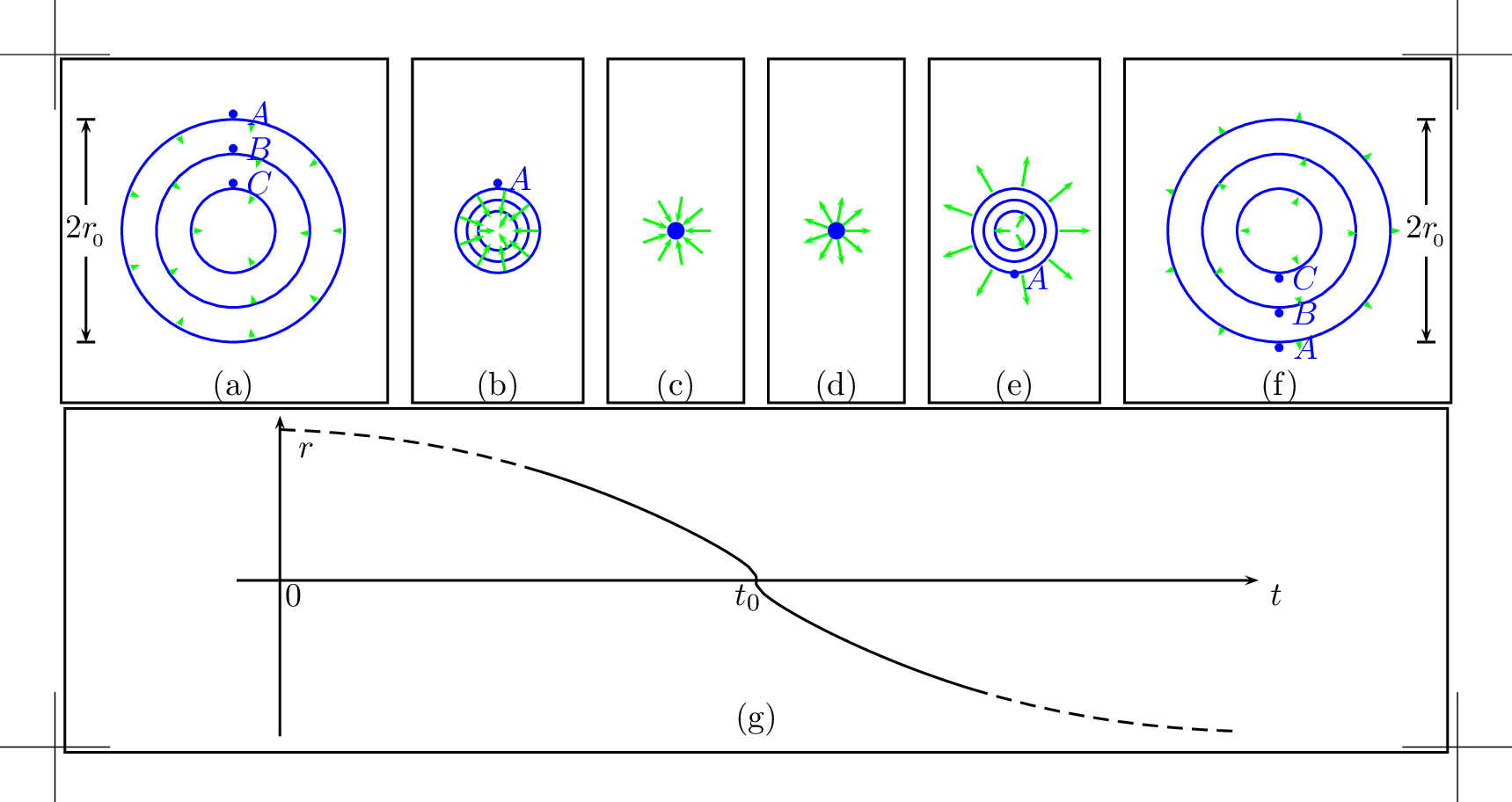}\end{center}
\caption{A complete evolution cycle of a collapsing star is from (a) to (b), then to (c), then to (d), then to (e), then to (f) and finally to (a) again. Depending on the initial conditions, evolutions from (a) to (b) and (e) to (f) could contain shell-crossing events. But the evolution from (c) to (d) contain shell-crossing events no matter how the initial conditions are. Subfigure (g) displays the radius evolution of the collapsing star, the fact that $t(<t_0)\rightarrow t_0$, $r=r_0(1-t/t_0)^\frac{2}{3}$ and as $t(>t_0)\rightarrow t_0$, $r=-r_0(t/t_0-1)^\frac{2}{3}$ follows from Newton mechanics $\ddot{r}=\frac{GM_\mathrm{tot}}{r^2}$ directly. General relativity would not change this fact qualitatively.}
\label{figBreathingBall}
\end{figure*}
In fact, the most important shell-crossing occurs on the central point. The crossing events there are unavoidable in both general relativity and newton mechanics and has no dependence on the stars' having a high density outer skin or not. They are results of momentum-energy conservation laws.  Consider an observer co-moving with the collapsing star, when it arrives near the central point
\beq{}
r=r_0\big(1-\frac{t}{t_0}\big)^\frac{2}{3},~\dot{r}=-\frac{2r_0}{3t_0}\big(1-\frac{t}{t_0}\big)^{-\frac{1}{3}},~t\rightarrow t_0-\epsilon
\eeq 
Its radial speed is divergent. So it cannot stop there immediately and have to shell-cross to the anti-direction
\beq{}
r=-r_0\big(1-\frac{t}{t_0}\big)^\frac{2}{3},~t\rightarrow t_0+\epsilon
\eeq
This means that static central singular point is not the final state of a collapsing star. The proper final state should be a periodically zero-cross breathing ball, see FIG.\ref{figBreathingBall} for pictures. In real collapsing star consisting of fermion particles, shell-crossing phenomenas are also unavoidable. Although near such crossing point, infinite pressures could appear due to Pauli-exclusion principles and lead to bounces of the crossing-wish shells, the bouncing itself could be thought as a shell-crossing when the identity principle is considered.
Obviously, this zero-cross breathing ball provides us a very smart way to resolve the central singularity of Schwarzshild black hole but successfully avoids contradictions with Penrose and Hawking's singularity theorem \cite{horowitz2016}. That is, central singularities indeed happen in finite times after the collapsing begins. However, it we continuously take photos inside the horizon of the system, what we get will be mostly of regular stars with various radius mass distributions instead a single point carries all the mass of system exclusively. We will show in the following that, this radial mass distribution provides just the physic basis for the micro-state of black holes, thus origins for the horizon entropy.

{\bfseries Inner structure of black holes, quantum description.} Obviously, if we can provide a quantum description for above classic pictures, our declaration that radial mass distributions inside the horizon of a collapsing star is just the micro-state of the equal mass black holes will be more believable to peoples. This is directive in the  quantum theory of gravitations originally proposed by B. S. DeWitt \cite{deWitt1967} and developed latter mainly in quantum cosmologies \cite{QCosmologyHalliwellLecture}. It is also applied to black holes exploration in references.\cite{QBHAllen1987,QBHFangLi1986,QBHLaflamme1987,QBHNambuSasaki1988,QBHNagai1989,QBHRodrigues1989}. However, none of these works tries to understand the micro-state of black holes by this method, although it is so natural and directive. To implement such descriptions, we firstly consider the 3+1 decomposed dynamics of matter and geometries inside the collapsing star
\beq{}
ds^2=-N^2dt^2+h^{-1}dr^2+r^2d\Omega_2^2,~h=1-\frac{2m(t,r)}{r}
\label{metric3plus1}
\eeq
\begin{gather}
\frac{S_L}{4\pi}=\!\!\int\!\!dtdrNh^{\m\frac{1}{2}}r^2\big[\frac{2m'}{r}
\!-\!\frac{\rho}{2}(\dot{x}\cdot\dot{x}+\!1)
\!-\!\frac{p}{2}(\dot{x}\cdot\dot{x}-\!1)\big]
\label{actionL3plus1}\\
\rule{30mm}{0pt}+\mathrm{local~total~derivative~terms~}
\nonumber
\end{gather}
where $\rho$, $p$ and $\dot{x}^\mu$ are the energy density, pressure and four velocity of fluid elements inside the collapsing star respectively. In the final equation of motion, normalisation $\dot{x}\cdot\dot{x}=-N^2\dot{x}^0\cdot\dot{x}^0+h^{-1}\dot{x}^r\cdot\dot{x}^r=-1$ should be set everywhere. And because $\dot{x}^r=-\frac{\dot{m}}{m'}$, $x^0=t$, $N^2=\frac{\dot{m}^2}{{m'}^2}h^{-1}+1$ follows from the 4-velocity's normalisation naturally. $N^2$ here being not independent variable has also counter sayings in cosmologies, where it is usually set as $N=1$ for the co-moving observers. So, in this 3+1 decomposed system \eqref{metric3plus1}-\eqref{actionL3plus1}, only $m(t,r)$ and $\rho$, $p$ are possible dynamic variables. Turning to the Hamiltonian language
\beq{}
P_m\equiv\frac{\delta S_L}{\delta\dot{m}(t,r)}
,~S_H=\int\!dr\,\dot{m}(t,r)P_m-S_L
\label{Pmdefinition}
\eeq
\begin{gather}
\frac{S_H}{4\pi}=\!\!\int\!\!dtdrNh^{\m\frac{1}{2}}r^2\big[-\frac{2m'}{r}
\!-\!\frac{\rho}{2}(N^2+h^{-1}\frac{\dot{m}^2}{{m'}^2}-\!1)
\label{actionH3plus1}\\
\rule{10mm}{0pt}+\!-\frac{p}{2}(\cdots)\big]-\mathrm{local~total~derivative~terms~}
\nonumber
\end{gather}
In the case $p=0$, hamiltonian constraint following from this action $\delta S_H/\delta N=0$ and the 4-velocity's normalisation will bring us expressions for $\rho$ completely the same as \eqref{rhoExpression}
\beq{}
\mathcal{H}(m,P_m)=h^{-\frac{1}{2}}r^2\big[-\frac{2m'}{r^2}-\rho\big(h^{-1}\frac{\dot{m}^2}{{m'}^2}+1\big)\big]=0
\label{Hconstraint}
\eeq
On the other hand, from the Hamilton-Jaccobi equation following from this action and the conservation law following from the vanishing of local total derivative terms, equations \eqref{eqConstraint} and \eqref{eqEvolution} could also be derived out routinely. This justifies the correctness of equations of motion written in the previous sections from the aspect of action principles.

Now, following ideas completely the same as quantum cosmologies \cite{QCosmologyHalliwellLecture,deWitt1967}, we consider $m(r)$ as a general coordinate and introduce a wave function $\Psi[m(r)]$ to denote the probability amplitude of the system with mass distributions $m(r)$. $\Psi[m(r)]$ satisfies the operator version of constraint \eqref{Hconstraint}, with $\dot{m}$ replaced by functions of $m$ and $P_m$, the latter by functional derivatives $\frac{-i\hbar\delta}{\delta m(r)}$
\beq{}
\big[8h^{\m\frac{3}{2}}r^2\!{m'}^{\m1}\rho-\frac{\hbar^2\delta^2}{\delta m(r)^2}
+4h^{\m\frac{3}{2}}r^4\!{m'}^{\m2}\rho^2\big]\Psi[m(r)]=0
\label{funcSchrodinger}
\eeq
This functional differential equation together with the following boundary condition [we use $H(x)$ denoting the usual Heaviside step function, so $H(x)=0~\mathrm{when}~x<0;~1~\mathrm{otherwise}$]
\beq{}
\Psi[m_hH(r-0)]\neq\infty,~
\Psi[m(r=r_h)<m_h]=0
\label{bndrSchrodinger}
\eeq
define a functional eigenvalue problem for $\Psi[m(r)]$. Similar to the usual eigenvalue problems in quantum mechanics, it can be imagined that only some special eigen-energy-densities $\{\rho_i(r),i=0,1,2\cdots\}$ could lead to normalisable wave-functional $\Psi_i[m(r)]$. Besides \eqref{funcSchrodinger} and \eqref{bndrSchrodinger}, the eigen-energy-density should also satisfy constraints
\beq{}
\int_0^{r_h}\!\rho_i(x)4\pi x^2dx\approx m_h
\label{consSchrodinger}
\eeq 
the approximation symbol here indicates our neglecting of the curved space fact in its written down.
With this final constraints, it's natural to conjecture that the index $i$ of eigenvalue/states has upper bound and the wave-functions $\{\Psi_i[m(r)],i=0,1,2\cdots,i_\mathrm{max}\}$ have one-to-one correspondence with the micro-state of the black holes in consideration.

To understand the fact that equations \eqref{funcSchrodinger}-\eqref{consSchrodinger} indeed defines the quantum micro-state of black holes, let us try to solve them by the following strategies, i) constraining $m(r)$ to the form $r^\xi$ so that $\frac{\delta}{\delta m}=(m\ln r)^{-1}\frac{\partial}{\partial\xi}$; ii) writing functionals $\Psi[m(r)]$ to usual functions $\Psi(\xi)$, thus changing the functional equation into a differential array
\begin{gather}
\forall r\in[0,r_h],~\big[\big(\frac{8mr}{\xi}r^2\!\rho
+\frac{4r^2}{\xi^2}r^4\!\rho^2\big)\big(1-\frac{2m}{r}\big)^{\m\frac{3}{2}}(\ln{\!r})^2
\label{normalEigenvalueproblem}
\\
\rule{20mm}{0pt}+\hbar^2\big(\ln{\!r\,}\partial_\xi-\partial^2_\xi\big)\big]\Psi(\xi)=0
,~m=\frac{r_h}{2}\big(\frac{r}{r_h}\big)^\xi
\nonumber
\\
\Psi(0)\neq\infty,~\Psi(\infty)=0,~\int_0^{r_h}\!\!\rho(x)4\pi x^2dx\approx m_h
\label{normalBoundcondition}
\end{gather}
This array contains infinite components, because its master equation need be satisfied as $r$ varies in the continuous range $[0,r_h]$. Operationally we can choose to let it be satisfied only on some discrete values of $r$. For example, in the 1$\ell_\mathrm{pl}$-sized black hole, we can choose such discrete points as $r=\frac{1}{6}$, $\frac{3}{6}$, $\frac{5}{6}$$\ell_\mathrm{pl}$ and specify the $r^2\rho(r)$ function by its values on three equal-width interval $(0,\frac{1}{3})$, $(\frac{1}{3},\frac{2}{3})$, $(\frac{2}{3},1)$. Assuming that mass/energy densities on each of these intervals be uniform, considering the total mass constraints \eqref{normalBoundcondition}, all possible $r^2\!\rho$ profiles could be listed as the following equation \eqref{oneMpltable}
\beq{}
\begin{tabular}{c|ccccc}
\diagbox{$r^2\!\rho\!\!\!$}{$\!\!\!r$}&${0\sim\frac{1}{3}}\atop{\frac{1}{6}}$&${\frac{1}{3}\sim\frac{2}{3}}\atop\frac{3}{6}$&${\frac{2}{3}\sim1}\atop\frac{5}{6}$&{m/e.dist.}&{ei.soln?}\\
\hline
1&3&0&0&\rhoONEdistribute{1}&$\star$\\
2&0&3&0&\rhoTWOdistribute{1}&$\circ$\\
3&0&0&3&\rhoTHREEdistribute{1}&$\circ$\\
4&2&1&0&\rhoFOURdistribute{1}&$\star$\\
5&1&2&0&\rhoFIVEdistribute{1}&$\circ$\\
6&0&2&1&\rhoSIXdistribute{1}&$\circ$\\
7&0&1&2&\rhoSEVENdistribute{1}&$\circ$\\
8&2&0&1&\rhoEIGHTdistribute{1}&$\star$\\
9&1&0&2&\rhoNINEdistribute{1}&$\circ$\\
0&1&1&1&\rhoTENdistribute{1}&$\star$
\end{tabular}
\label{oneMpltable}
\eeq
\begin{equation}
\begin{tabular}{c|cccccc}
\diagbox{$r^2\!\rho\!\!\!$}{$\!\!\!r$}&${0\sim\frac{1}{3}}\atop{\frac{1}{6}}$&${\frac{1}{3}\sim\frac{2}{3}}\atop\frac{3}{6}$&${\frac{2}{3}\sim\frac{3}{3}}\atop\frac{5}{6}$&${\frac{3}{3}\sim\frac{4}{3}}\atop\frac{7}{6}$&${\frac{4}{3}\sim\frac{5}{3}}\atop\frac{9}{6}$&${\frac{5}{3}\sim\frac{6}{3}}\atop\frac{11}{6}$\\
\hline
1&6&0&0&0&0&0\\
2&0&6&0&0&0&0\\
\vdots&\vdots&\vdots&\vdots&\vdots&\vdots&\vdots\\
6&0&0&0&0&0&6\\
7&5&1&0&0&0&0\\
8&0&5&1&0&0&0\\
\vdots&\vdots&\vdots&\vdots&\vdots&\vdots&\vdots\\
11&0&0&0&0&5&1\\
12&1&5&0&0&0&0\\
13&0&1&5&0&0&0\\
\vdots&\vdots&\vdots&\vdots&\vdots&\vdots&\vdots\\
462&1&1&1&1&1&1\\
\end{tabular}
\label{twoMpltable}
\end{equation}
\begin{figure}[h]
\begin{center}\includegraphics[scale=0.5]{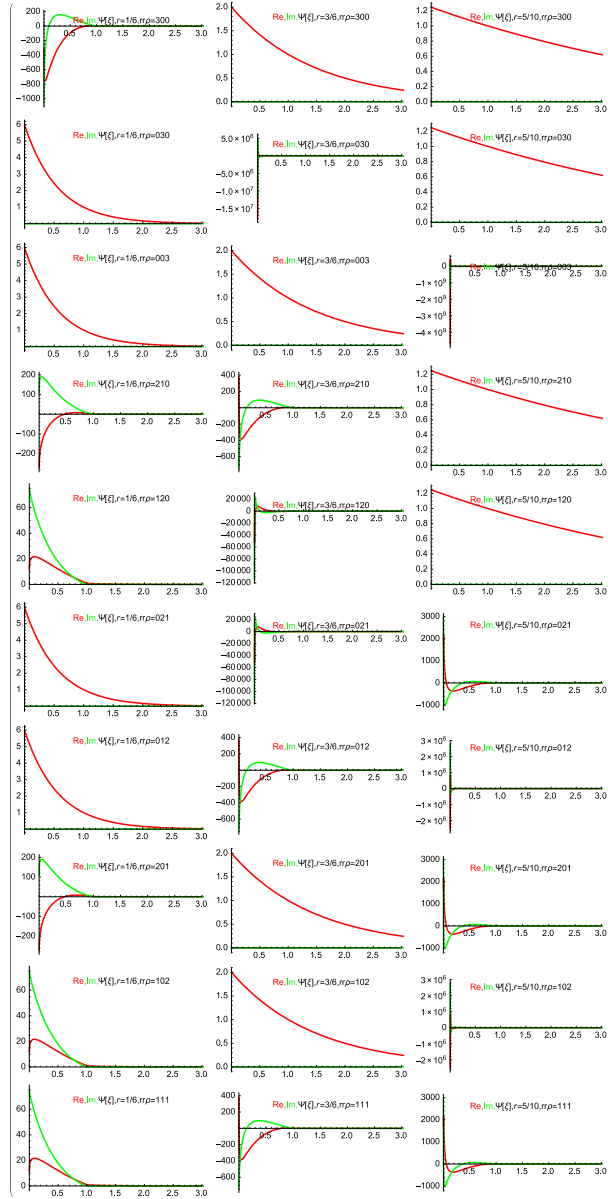}\end{center}
\caption{Numerical solutions to the differential equation \eqref{normalEigenvalueproblem} with conditions $\Psi(\xi=1)=1$, $\Psi(\xi=\infty)=0$ for $10$ possible $r^2\rho$ profiles listed in \eqref{oneMpltable} on positions $r=\frac{1}{6}$, $\frac{3}{6}$, $\frac{5}{6}\ell_\mathrm{pl}$. Good eigen-wave-function should be such ones that i) $\Psi(0)\neq\infty$, ii) normalisable and iii) $r$-independent as possible as can be.}
\label{figPsi1pl}
\end{figure}
For each of these $r^2\rho$ profiles we solve equations \eqref{normalEigenvalueproblem}[all solutions are normalised to $\Psi(\xi=1)=1$ and $\Psi(\xi=\infty)=0$] on the interval centrals $r=\frac{1}{6}$, $\frac{3}{6}$, $\frac{5}{6}$. The results are displayed in FIG.\ref{figPsi1pl}, from which it can be easily see that, if we insist a good eigen-energy-density need satisfy i) $\Psi(0)\neq\infty$, ii) normalisable and iii) being $r$-independent exactly, then none of the ten mass/energy distributions listed in \eqref{oneMpltable} is a good one. However, if we discretised function $r^2\rho(r)$ on more finer grids, we can obtain eigen-energy-density profiles more close to these judgements. On the 3-interval discretising level, the 1st, 4th, 8th and 10th profle in \eqref{oneMpltable} could be looked as good ones. The key point here is that, the system configuration with all mass/energy concentrating on the central point --- the 1st one --- is not the unique good eigen-energy-distribution. Instead, the good eigen-distribution is $4\sim e^{1^2}$-times degenerate.

Further, if we consider the 2$\ell_\mathrm{pl}$-sized black hole, we will find that if the same precision as 1$\ell_\mathrm{pl}$-sized black hole is wished, then the discretising of function $r^2\rho(r)$ should be on $6$ equal-length interval, the number of all possible profiles adds up to 462, some of them are listed in equation \eqref{twoMpltable} explicitly. Similar to 1$\ell_\mathrm{pl}$-sized black holes, we find that not all these mass/energy profiles are equally good eigen-energy-densities that makes the quantum wave-function i) $\Psi(0)\neq\infty$, ii) normalisable and iii) $r$-independent to the highest degree. We find that, the good eigen-energy-distribution scheme is approximately $55\sim e^{2^2}$-times degenerate. Now, if we want to use this same idea as in 1 and 2$\ell_\mathrm{pl}$-sized black holes to more larger ones, we will need to numerically solve exponentially-many schr\"odinger equation to find the good eigen-energy-density solutions, which is obviously impossible operationally. However, explorations in the small black hole examples indeed provides us supporting evidence that eigen-energy-densities defined by eqs.\eqref{funcSchrodinger}-\eqref{consSchrodinger} have one-to-one correspondence with the micro-state of black holes. We introduce in the following an approximate method for the number counting of eigen-states of large black holes by the so called correspondence principles \cite{QMprinciple}.

{\it\bfseries The micro-states' number counting and horizon entropies.} As is well know, in collapsing stars corresponding to very large black holes, the average density of the system is very small $\rho_\mathrm{av}\approx M/(2GM)^3$. According to Newtonian mechanics, the collapsing speed of these large stars is correspondingly very small due to the fact that the collapsing time square $t^2\propto 1/G\rho_\mathrm{av}$. According to equation \eqref{rhoExpression}, the local energy density of the system is thus approximately $\rho\approx\frac{m'}{4\pi r^2}$, which is just the density definition is conventional Newton mechanics. This means that for large black holes, the number counting of proper eigen-energy-density profiles $\rho(r)$ could be replaced by the number counting of mass function $m(r)$ directly. On the other hand, our numeric examples in the second section of work also tell us that for the initially non-singular collapsing stars, the horizon always forms earlier than central singularities. So matter distributions an infinitesimal time before or after the horizon forms could be looked as ideal proxies of the system's quantum states. Obviously, the idea here is very similar to the correspondence principle firstly introduced by N. Bohr in early quantum mechanics in establishing relations between the quantum wave function and classic orbits of electrons in atoms \cite{QMprinciple}. The key question here is, how to make the continuous mass function $m(r)$ become discrete object thus count them one by one.

\begin{figure}[h]
\begin{center}\includegraphics[scale=0.6]{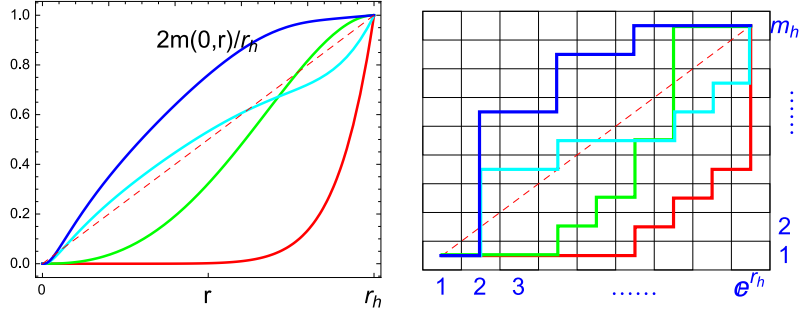}\end{center}
\caption{The left hand side displays four mass distribution ways inside a collapsing star an infinitesimal time before/after it collapses into the horizon. Along the diagonal line $2m(0,r)=r$. The right hand side is a discrete representation of the left. Each continuous distribution way corresponds to a regular Young diagram of $m_\mathrm{h}/m_\mathrm{pl}$ row and $1\sim e^{r_h/\ell_\mathrm{pl}}$ column.}
\label{figEntropy}
\end{figure}
Our idea is, introduce an ``distribution quantisation'' assumption so that any two collapsing stars with equal total mass but different radial distributions by 1-$m_\mathrm{pl}$ mass shells of radius $r\leqslant r_h$ and concentric with the parental star could be identified as distinguishable quantum states of the corresponding black hole. Similar to Bohr-Sommerfield's orbital quantisation condition leading to discretised orbit for electrons in atoms, our distribution quantisation condition will lead to discretised radial mass distributing ways. Referring to FIG. \ref{figEntropy}, our quantisation condition requires the vertical line of the distribution function space be discretised by $m_\mathrm{pl}$, while to distinguish those distributions differ by only 1-$m_\mathrm{pl}$ mass shells near the horizon edges, the horizontal line must be discretised exponentially, which arises from the logrithmic divergence of mass functions there, see equation \eqref{mfLGsingular}.

Obviously, in the discretised function space, each radial mass distribution corresponds to a regular Young diagram of $m_\mathrm{h}/m_\mathrm{pl}$ row and $1\sim e^{r_h/r_\mathrm{pl}}$ column. So the total number of such distributions is $W=(e^{\frac{r_h}{r_\mathrm{pl}}})^{\frac{m_h}{m_\mathrm{pl}}}$. Since $r_hm_h\propto r_h^{n-1}\propto$ horizon area in $n+1$ dimensional space-time, this implies that, associating with every Schwarzschild black hole of horizon area $A$, is a micro-canonical ensemble of $e^{A/A_\mathrm{pl}}$ collapsing stars, all with the same mass and surface area but each with different inner mass distributions.  According to definitions, the entropy of such black holes reads
\beq{}
S=k_B\log W=\frac{A}{A_\mathrm{pl}}
\label{formulaEntropy}
\eeq
This is nothing but the Bekenstein-Hawking formula \cite{Wald1999} up to a numeric factor of order $1$. It is worth to emphasise that, our micro-states counting involves only initial distributions $m(0,r)$ instead of $m(0,r)$ and $\dot{m}(0,r)$ simultaneously. This is because, Einstein equation gives two superficially redundant components controlling the evolution $m(\tau,r)$. Given initial distributions, component \eqref{eqConstraint} will fix the speed $\dot{m}(0,r)$ while \eqref{eqEvolution} will yield dynamical evolutions $\ddot{m}(0,r)$ to the next epoch. This is remarkably different from other dynamical systems which are controlled by only one differential equation and is the key reason for area laws. 

It should be emphasized that the micro-state we counted here is not local motion modes of particles inside the black hole. Such local degrees of freedom contribute to the micro-state of the system only of order $e^{n(\mathrm{i.e.~particle~number})}\sim e^{m_h}\sim e^{r_h}$, which is obviously negligible relative to the radial mass distribution modes $e^{r_h^{n-1}}$ in $n+1$ dimensions. The micro-state we counted here is non-local collective motion of matter contents inside the black hole, they are essentially geometrical degrees of freedom because their form uniquely determines inner geometries of the system. In the series of works \cite{nonlocal2008,nonlocal2009,nonlocal2014,nonlocal2015}, Stojkovic et al provides many concrete evidences that, non-locality plays key roles in both the central singularities resolution and the Hawking radiations's unitarity recovering.

{\it\bfseries Solutions to the information missing puzzle.} The above pictures for the micro-state of black holes implies a direct method to resolve the information missing puzzle. To see this more explicitly, we rewrite equation \eqref{eqConstraint} in first order forms, but in this time understand the mass function $m(\tau,r)$ as the inner mass distribution of the black holes in special states,
\begin{gather}
\dot{m}^2(\tau,r)={m'}^{2}(\tau,r)\cdot\exp\big[\!\!\int_0^r\!\!\frac{4m(\tau,x)dx}{x(x-2m)}\big]
\times\\
\int_0^r2m(\tau,x)x^{-2}\exp\!\big[\!-\!\int_0^x\!\frac{4m(\tau,y)dy}{y(y-2m)}\big]dx
\nonumber
\end{gather}
Obviously, each micro-state of the black hole has its own characteristic inner mass distribution, thus characteristic speed of total mass variation $\dot{m}(\tau,r=r_\mathrm{edge})$ when they evaporate/accretes. By recording this speed of mass/size variation, we could reproduce all the information related with its inner mass distributions. So there are no information missing puzzles related with the Hawking radiation! This almost classic general relativistic resolution of information missing puzzles is possible because, in both Hawking's original calculation \cite{hawking1976} and the latter advanced version of F. Wilcek \cite{Wilczek2006}, the  background black holes are assumed to have fixed horizon sizes, thus imposes no effects on the evaporation speed. These calculations could provide dynamic mechanisms by which particles escape from the horizon. But they have no chances to catch kinematics of the background black hole's size variation. It is just this kinematics that carries away the missed information. 

The area law and micro-interpretation for the black holes' entropy lie on centers in string theory and loop gravity's achievements. However, none of them, including the recent interpretations of Hawking, Perry and Strominger, is verifiable experimentally. As comparisons, our interpretations in this work is dis/verifiable observationally. Since the information of black holes in our interpretation is identified with radial mass distributions of the corresponding collapsing stars, it could be extracted or released through certain classic process. For example, in astrophysical events such as binary black hole's mergering \cite{gw150914,gw151226}, signals other than gravitational waves such as gamma ray bursts could be produced when matters going from one hole to the other. Reference \cite{grb150914obsv,grb150914expl,gw150914otherInterpretation} may have given us such evidences already. Even when gamma rays are non-available, the form of gravitational waves would have different shapes when produced from binaries with different inner distributions. With the development of gravitational wave and gamma ray astronomies, this verification may already be at our technique abilities.  

{\it\bfseries Conclusion.} We provide in this work the most general dynamic inner metric of collapsing stars with horizon and non-trivial radial mass distributions. We find that near the central singular point, shell-crossing phenomena is unavoidable and static central singularities are not the final state of all such collapsing stars. Instead, their final state is something we called zero-cross breathing balls. This naturally resolves the central singularity of Schwarzschild black holes but avoids contradiction with Penrose and Hawking's singularity theorem. If we take photos for these breathing ball in their horizon continuously, then what we get will be mostly of collapsing star with various regular radial mass distribution instead of singular points concentrating all masses of the system exclusively. The radial mass distribution here is nothing but micro-states that lead to horizon entropies for the black holes with equal masses. Non-radial local random motions of particles inside an externally-looking spherical symmetric collapsing star are although possible, due to the fact that the particle number inside the horizon linearly depends on the mass thus on the horizon size of the black hole in consideration, they contribute to the entropy of the system only of $\mathcal{O}[r_h]$, obviously negligible relative to the horizon entropy $\mathcal{O}[r_h^{n-1}]$ in $n+1$ dimensional space-times. 

We then enhance the above classic picture in quantum formulations of the 3+1 decomposed general relativity further. We find that the micro-state of the zero-cross breathing ball is defined by the eigen-energy-density solution of a functional Schr\"odinger equation.  For 1 and 2$\ell_\mathrm{pl}$ sized this kind of ball, we provide numeric evidence that the eigen-solution is about $e^{1^2}$ and  $e^{2^2}$times degenerate. While for large this kind of ball, by assuming that any two distributions with equal total mass but different radial profiles by any 1$m_{pl}$-weighted mass-shells correspond to distinguishable quantum states, we show that the degeneracy is approximately of $e^{r_h^2/r^2_\mathrm{pl}}$ order. Since each of these degenerating ball has special de-horizon/expansion speed determined by its radial mass distribution and could be measured as its identifying accordance, no information will be missed during a black hole's evaporation. We thus provide not only a microscopic interpretation for the horizon entropy of black holes, but also a concrete resolution to the information missing puzzle involved in their Hawking radiations.

Obviously, it is a progress to translate the question of micro-state definition and number-counting related with the black hole entropy into solution's searching of a functional eigen-value-problem. However, since we find no exact solutions, we still have distances to the final solutions to these questions exactly. So, as the first suggestion for future works, we think that, finding highly-effective numeric algorithm or systematic approximation shcem to solve equations \eqref{funcSchrodinger}-\eqref{consSchrodinger} maybe the most important work to do. Our second suggestion is,  since our physic picture implies that black holes are nothing but micro-ensemble of collapsing stars with the same mass but different radial mass distributions, it is very interesting to quantitatively investigate differences between the shape of gravitational waves produced in the mergering of binary black holes with different inner mass distribution. Such investigations are still absent on the market \cite{GWbinary2005,GWbinary200601,GWbinary200602} and will be very useful for the future observational dis/verification of our pictures. Thirdly, considering the non-local essence of the micro-state corresponding to the horizon entropy, it is very important to investigate the relation-ship between our definitions through functional Schrodinger equation and that through quantum entanglements \cite{Srednicki1993,RyuTakayanagi2006}. Other prospects such as generalising our discussions to some more general black holes or de Sitter space-time itself may also  be possible and interesting.

\section*{Acknowledgements}
The author thanks very much to J.-b. Wu, Y. H. Gao, G. Horowitz, D. Gross and E. Witten (during the 2016 international string theory conference) for their valuable comments and suggestions in the preparation and revisions of this work. This work is supported by Beijing Municipal Natural Science Foundation, Grant No. Z2006015201001 and partly by the Open Project Program of State Key Laboratory of Theoretical Physics, Institute of Theoretical Physics, Chinese Academy of Sciences, China.

\end{document}